\documentclass[10pt]{article}
\input epsf

\begin{document}

\title{The statistical nature of the second order corrections to the thermal SZE}
\small{\author{A. Sandoval-Villalbazo$^a$ and L.S. Garc\'{\i}a-Col\'{\i}n$^{b,\,c}$ \\
$^a$ Departamento de F\'{\i}sica y Matem\'{a}ticas, Universidad Iberoamericana \\
Lomas de Santa Fe 01210 M\'{e}xico D.F., M\'{e}xico \\
E-Mail: alfredo.sandoval@uia.mx \\
$^b$ Departamento de F\'{\i}sica, Universidad Aut\'{o}noma Metropolitana \\
M\'{e}xico D.F., 09340 M\'{e}xico \\
$^c$ El Colegio Nacional, Centro Hist\'{o}rico 06020 \\
M\'{e}xico D.F., M\'{e}xico \\
E-Mail: lgcs@xanum.uam.mx}} \maketitle

\bigskip
\begin{abstract}
{\small {This paper shows that the accepted expressions for the
second order corrections in the parameter $z$ to the thermal
Sunyaev-Zel'dovich effect can be accurately reproduced by a simple
convolution integral approach. This representation allows to
separate the second order SZE corrections  into two type of
components. One associated to a single line broadening, directly
related to the even derivative terms present in the distortion
intensity curve, while the other is related to a frequency shift,
which is in turn related to the first derivative term. }}
\end{abstract}

The thermal Sunyaev-Zel'dovich effect (SZE) \cite{SZ1} \cite{SZ2}
arises from the frequency shift of CMB photons that are scattered
by a hot electron gas. It has been detected in observations of
some rich, X-ray luminous clusters. The SZE can be applied to the
determination of cosmological parameters \cite{Birk1}
\cite{Steen}.

The distorted spectra arising from Compton scattering can be
expressed as a power series containing terms which are even powers
of the frequency $\nu$, except for a first  order term in $\nu$
that is associated to the photon diffusion approximation
\cite{Sazonov} \cite{Stebbins} \cite{Itoh} \cite{Challinor}. The
parameter $z=\frac{k T_{e}}{m c^{2}}$ used as a discriminant to
indicate when the relativistic corrections are important is
usually kept up to second order effects.

In this paper we show that the scattering law approach to the SZE
\cite{nos1}  is able to reproduce the distortion curve contained
in Refs. \cite{Sazonov} \cite{Stebbins} \cite{Itoh}
\cite{Challinor} \cite{Rephaeli}  providing new insight of the
physics inherent in distortion spectrum. The basic idea is
simple. The distorted spectrum can be obtained by performing a
convolution integral between the CMB blackbody curve and a kernel
corresponding to a single line profile, slightly broadened by the
interaction between photons and the thermal electrons present in
the gas. For the case of distortion curves linear in $z$ and in
the optical depth $\tau$, it has already been shown that the line
broadening corresponding to a Maxwellian reproduces the Kompaneets
equation \cite{nos3}.

To obtain the widely accepted spectra from a scattering law, in
the relativistic case,  we have chosen to start from a slight
modification of a kernel recently proposed and perform explicitly
the convolution integral. Having obtained the distortion
expression we shall compare it with the the distortion curve
derived from references \cite{Sazonov} \cite{Stebbins} \cite{Itoh}
\cite{Challinor} \cite{Rephaeli}. A brief discussion is included
at the end of this note.

The scattering law approach to CMB distortions is based upon the
fact that the distorted radiation spectrum which originates from
photon scattering in dilute medium is given by:

\begin{equation}
I(\nu) = \int_{0}^{\infty} I_{o} (\bar{\nu}) G_{s} (\bar{\nu},\nu)
d \bar{\nu} \label{uno}
\end{equation}

Here, $I(\nu)$  is the scattered radiation off the plasma, $
I_{o}(\nu)=\frac{2 h \nu^{3}}{c^{2}}(exp(\frac{h \nu}{k
T_{R}})-1)^{-1}$ the undistorted spectrum, where $T_{R}$ is the
CMB temperature,  and $G_{s} (\bar{\nu},\nu)$ the scattering law.

For the thermal non-relativistic effect we have successfully
reproduced Kompaneets equation with a kernel  $ G_{s}
(\bar{\nu},\nu)$ which is given by \cite{nos2}

\begin{equation}
G_{s} (\bar{\nu},\nu) =(1-\tau) \delta (\bar{\nu}-\nu) +  \tau
G(\bar{\nu},\nu) \label{dos}
\end{equation}
where
\begin{equation}
G(\bar{\nu},\nu)=\frac{1}{\sqrt{\pi } W(\nu)}
e^{-(\frac{\bar{\nu}-(1-2z)\nu}{W(\nu)})^{2}} . \label{tres}
\end{equation}

Here, $m_{e}$ and $T_{e}$ are the mass and temperature of an
electron and the electron gas, respectively, while $W^{2}(\nu)=4
\frac{k T_{e}} {m_{e} c^{2}} \nu^{2}=4 z \nu^{2}$ is, obviously,
the square of the width of the spectral line at frequency $\nu$.

We now introduce a modified kernel given by:
\begin{equation}
G_{s}(\bar{\nu},\nu) = (1-2\tau) \delta(\bar{\nu}-\nu)
+\frac{\tau}{\sqrt{\pi} W(\nu)}
e^{-(\frac{\bar{\nu}-\nu}{W(\nu)})^{2}} +\tau
\delta(\bar{\nu}-(1-2z)\nu) \label{cuatro}
\end{equation}

The proposed kernel, Eq. (\ref{cuatro}), simply implies that the
single line profile is a purely Gaussian function peaked at
$\bar{\nu}=\nu$ \cite{Sun1} and that the central limit theorem
governs the scattering of photons by electrons \cite{Ned}. Also, a
systematic effect, associated to a $2 z \nu$ frequency shift is
present consistently with the diffusion approximation. Both
effects are small and are assumed additive. The Gaussian term in
the corresponding integral (\ref{uno}) becomes even in the
variable
\begin{equation}
\alpha= \frac{\bar{\nu}-\nu}{W(\nu)}\label{cinco}
\end{equation}
leading to the expression, up to second order in $z$:
\begin{equation}
I(\nu)=I_{o}(\nu)-2\tau I_{o}(\nu)+
\frac{\tau}{\sqrt{\pi}W(\nu)}\int_{-\infty}^{\infty}
I_{o}(\nu+\Delta \nu (\alpha))e^{-\alpha^{2}} d\alpha +\tau
I_{o}(\nu-2z \nu)\label{seis}
\end{equation}
where $\Delta \nu(\alpha)=W(\nu) \alpha $.

The expression $I_{o}(\nu+\Delta \nu (\alpha))$ can be trivially
expanded up to arbitrary order in $z$, but for the purposes of
this note we will only consider the second order corrections. The
resulting distortion curve reads:
\begin{equation}
\begin{array}{c}
\frac{\Delta I}{\tau }=-2z \nu \frac{\partial I_{o}}{
\partial \nu }+(z+\frac{1}{2}z^{2})\,\nu ^{2}\frac{\partial ^{2}I_{o}}{
\partial \nu ^{2}} \\
+\frac{1}{2}z^{2} \nu^{4} \frac{\partial ^{4}I_{o}}{\partial \nu
^{4}}
\end{array}
\label{cincob}
\end{equation}

In obtaining Eq. (\ref{cincob}), the last term of Eq. (\ref{seis})
was also expanded up to second order in $z$. A careful examination
of the result corresponding to the relativistic thermal SZE
derived by Sazonov and Sunyaev for a static cluster ($V=0$) leads
to the widely accepted expression \cite{Sazonov}:

\begin{equation}
\begin{array}{c}
\frac{\Delta I_{SS}}{\tau }=(-2\,z+\frac{17}{5}z^{2})\,\nu
\frac{\partial I_{o}}{
\partial \nu }+(\,z-\frac{17}{10}z^{2})\,\nu ^{2}\frac{\partial ^{2}I_{o}}{
\partial \nu ^{2}} \\
+\frac{7}{10}z^{2} \nu^{4} \frac{\partial ^{4}I_{o}}{\partial \nu
^{4}}
\end{array}
\label{cincoc}
\end{equation}

Fig. (1) shows a direct comparison between the distortion curve
that arises from Eq. (\ref{cincoc}) and the one arising form the
convolution kernel (\ref{cuatro}), for a temperature of $k
T_{e}=10$ KeV. The convolution kernel (\ref{cuatro}) reproduces
the main features of the relativistic thermal SZE. Shimon and
Rephaeli \cite{Shimon} have remarked  that all the current
independent approaches to the relativistic thermal SZE are
compatible with Eq. (\ref{cincoc}). In our Eq. (\ref{cincob}), the
$ z^{2 }\nu^{4}$ term arises from the purely
\emph{non-relativistic} line broadening $(W (\nu))^{4}$ yielding a
$\frac{5}{10}$ factor instead of the $\frac{7}{10}$ factor
included in Eq. (\ref{cincob}).

\begin{figure}
\epsfxsize=3.4in \epsfysize=2.6in
\epsffile{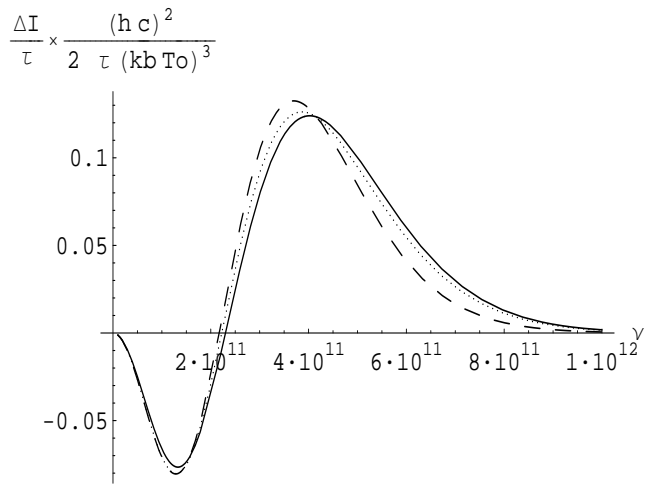}\vspace{0.5cm} \caption {\small{CMB
distortion for the case of Eq. (7) (dotted line)  compared with
the relativistic curve, Eq. (8) (solid line), $k T_{e}=10 KeV$.
The long dashed line has been added as a reference and corresponds
to the Kompaneets approximation. The frequency is given in Hz and
$\frac{\Delta I}{\tau}$ is given in units of $\frac{2 (k
T_{R})^{3}}{(h c)^{2}}$.} }
\end{figure}
\vspace{0.5cm}

We wish to emphasize on the fact that the high frequency behavior
of the distorted spectrum has been obtained using a superposition
of the distortion effects associated to the Gaussian Kernel and
the $-2z \nu$ frequency shift established in Ref. \cite{nos1}. It
is also interesting to notice that the suppression of the
$\frac{17}{5} z^{2}$ and $ \frac{17}{10} z^{2}$ terms in Eq.
(\ref{cincoc}) would not alter in a significant way the spectrum
up to temperatures of $k T_{e}=10 KeV$. This means that the
relativistic corrections to the thermal SZE may be interpreted as
small second order modifications to the shift and width parameters
included in Eq. (\ref{cuatro}).

The inclusion of the shift in the delta function in the
convolution Kernel preserves the even parity in the higher order
derivative terms of the distortion curve up to second order in
$z$. A possible alternative that would yield both even and odd
higher order derivative terms has been discussed elsewhere
\cite{black}.

Similar techniques as the one  applied here to CMB physics have
been discussed in other contexts \cite{Fixsen}, and are specially
attractive to derive a first analysis of the nature of the
distortions that arise from the interaction of photons and dilute
systems.

The authors wish to thank Prof. Roy Maartens (ICG, Portsmouth) for
his valuable comments.

This work has been supported by CONACyT (M\'{e}xico), project
$41081-F$.

\end{document}